\documentclass[aps,reprint,prl,groupedaddress]{revtex4-2}
\usepackage{xcolor}
\usepackage{float}
\usepackage{amsmath,amsfonts,amssymb,epsfig,graphicx}
\usepackage{slashed} 
\usepackage{hyperref}
\usepackage[normalem]{ulem}
\usepackage{lineno}
\usepackage{physics}

\begin{document}

\title{Suitability Studies of Exotic Muon Decay at the High Intensity heavy-ion Accelerator Facility}

\author{Lingzhi \surname{Dong}$^{1}$}
\email[]{donglingzhi0620@stu.pku.edu.cn}

\author{Jinning \surname{Li$^{1}$}}
\email[]{lijinning@stu.pku.edu.cn}

\author{Leyun \surname{Gao}$^{1}$}

\author{Cheng-en \surname{Liu}$^{1}$}

\author{Yu \surname{Xu}$^{3,4}$}

\author{Xueheng \surname{Zhang}$^{2,3,4}$}

\author{Liangwen \surname{Chen}$^{2,3,4}$}

\author{Qite \surname{Li}$^{1}$}
\email[Contact author: ]{liqt@pku.edu.cn}

\author{Chen \surname{Zhou}$^{1}$}
\email[Contact author: ]{czhouphy@pku.edu.cn}

\author{Qiang \surname{Li}$^{1}$}
\email[Contact author: ]{qliphy0@pku.edu.cn}

\author{Zhiyu \surname{Sun}$^{2,3,4}$}

\affiliation{$^{1}$School of Physics and State Key Laboratory of Nuclear Physics and Technology, Peking University, Beijing, 100871, China}

\affiliation{$^{2}$School of Nuclear Science and Technology, University of Chinese Academy of Sciences, Beijing 100049, China}

\affiliation{$^{3}$Advanced Energy Science and Technology Guangdong Laboratory, Huizhou 516000, China}

\affiliation{$^{4}$Institute of Modern Physics, Chinese Academy of Sciences, Lanzhou 730000, China}

\begin{abstract}

We present a detailed feasibility study of exotic muon decay searches using the muon beam at the High Intensity heavy-ion Accelerator Facility (HIAF). For the decay channel \(\mu\to eX^0\), where \(X^0\) denotes a new boson featuring LFV couplings, the produced electron carries very low energy when the mass of $X^0$ approaches the muon mass, making detection challenging. Nevertheless, the high‑energy beam at HIAF can boost these electrons up to a measurable energy range. The detector system proposed here comprises front-end scintillators for particle identification (PID) and a main detector consisting of an RPC stack to track decay products and a scintillator for energy measurement. Leveraging the kinematic properties of low momenta and small emission angles of signal electrons, we employ optimized momentum and angular acceptance cuts on the RPC detector to efficiently suppress background events. With this strategy, we obtain an upper limit of $10^{-5}$ on the branching ratio of the exotic muon decay at 95\% confidence level, achieving state-of-the-art sensitivity.

\end{abstract}

\maketitle

\section{Introduction} 

The Standard Model (SM) of particle physics stands as the most successful theoretical framework for describing elementary particles and their interactions to date, having withstood stringent precision tests across numerous experiments over the past decades. Within the lepton sector, the three charged leptons — the electron ($e$), muon ($\mu$), and tau ($\tau$) — carry distinct flavor quantum numbers. At tree level, flavor-changing neutral current (FCNC) processes are forbidden, rendering charged lepton flavor an exactly conserved quantum number. Nevertheless, the experimental discovery of neutrino oscillations\cite{Super-Kamiokande:1998kpq,SNO:2001kpb} provides the first direct evidence that lepton flavor is not conserved in the neutral lepton sector, a phenomenon known as lepton flavor violation (LFV).

This observation challenges the original formulation of the SM and strongly suggests that charged lepton flavor violation (CLFV) processes may also exist in the charged lepton sector. Muons, being long-lived leptons with well-understood decay channels, serve as ideal probes for investigating CLFV processes\cite{LeeRoberts:2007gf}. Numerous experimental searches have targeted theoretically predicted muon decay channels such as $\mu\to e\gamma$, $\mu\to3e$ and $\mu N\to eN$\cite{Perrevoort:2023jry,Marciano:2008zz,Perez:2021gnr,SINDRUMII:2006dvw}; observation of any such CLFV signal would constitute evidence for physics beyond the Standard Model. Furthermore, the process $\mu\to eX^0$ was proposed\cite{PhysRevLett.57.2787} as a channel to search for the massive neutral boson $X^0$ featuring LFV couplings\cite{Heeck:2017xmg} such as Goldstone boson, familon, majoron, etc. The mass of $X^0$ is constrained by $m_{X^0} \leq m_\mu - m_e$, and in the muon rest frame the kinematics satisfy:

\begin{equation}
    E_{e}^*=\frac{m_\mu^2 + m_{e}^2 - m_{X^0}^2}{2m_\mu}, \ \ E_{X^0}^*=m_{\mu}-E_e^*.
    \label{eq1}
\end{equation}

\begin{figure}
    \centering
    \includegraphics[width=0.9\linewidth]{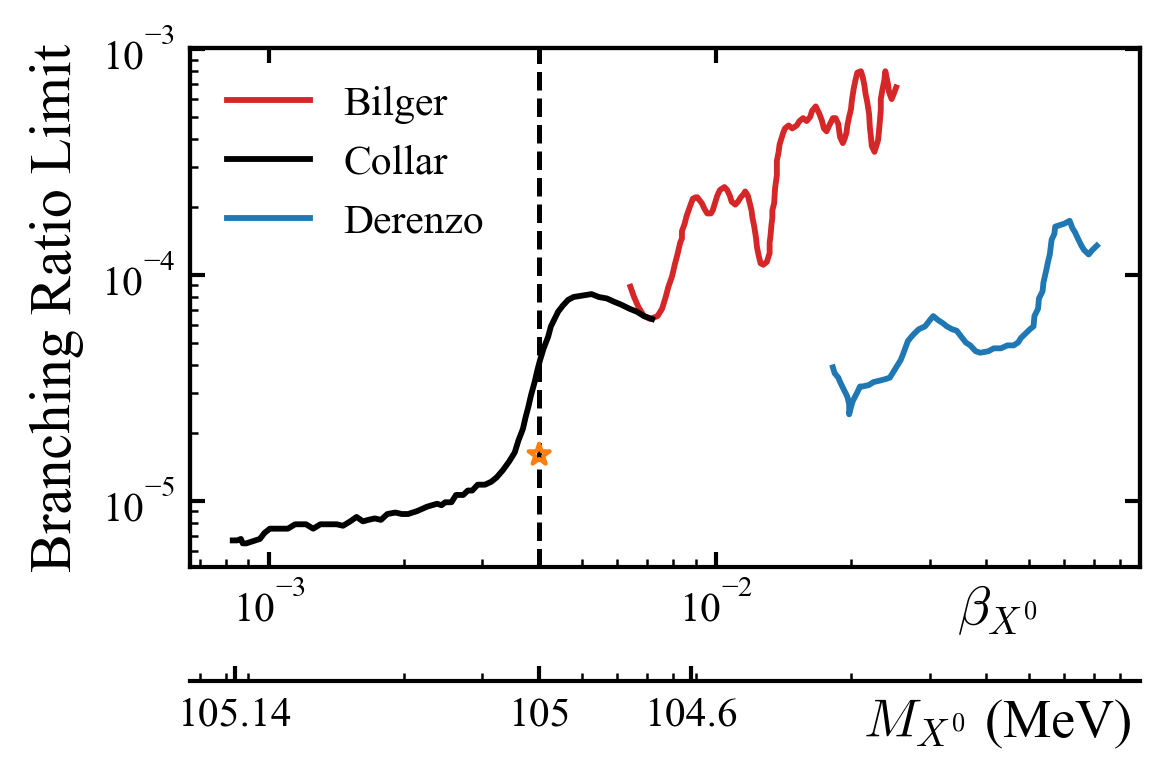}
    \caption{The upper limits on the branching ratio of the $\mu\to eX^0$ decay reported by existing studies under different $m_{X^0}$ hypotheses. The figure is reproduced from \cite{Collar:2023xla}. Results shown in the figure correspond to \cite{Collar:2023xla,Derenzo:1969za,Bilger:1998rp} in the order given in the legend. The yellow satr shows our results of this work, where We adopt a representative mass point as our benchmark.}
    \label{fig:introduction_results}
\end{figure}

Existing experiments have searched for this boson across a broad mass range\cite{Bilger:1998rp,Collar:2023xla,Derenzo:1969za,Bryman:1986wn,TWIST:2014ymv,PIENU:2020loi}; some of these studies also focus on the heavy‑mass \(X^0\), and these results are presented in Fig.~\ref{fig:introduction_results}. By comparison, our focus lies on \(X^0\) near the muon mass. From Eq.~\ref{eq1}, the decay electrons are produced with very low energies and are hardly detectable with current approaches. Collar et al. had addressed this issue by utilizing the low-energy positive muon beam from the M20 beamline at the TRIUMF facility in Canada, combined with high-resolution n-type low-energy germanium detectors. This approach pushes the upper limit for heavy \(X^0\) searches down to \(10^{-5}\)\cite{Collar:2023xla}. As an alternative strategy, high-energy muon beams can provide a substantial Lorentz boost, elevating intrinsically low-energy decay electrons into a detectable energy regime. This is the core concept adopted in the present work.

The High Intensity heavy-ion Accelerator Facility (HIAF)\cite{Xu:2025spd} is the heavy-ion accelerator facility that delivers the highest intensities in the world for low-energy continuous beams and high-energy pulsed beams. It can provide high-energy GeV muon beams with a momentum spread of 2\% and beam intensity between $10^5$ and $10^6$ particles per second, making it an ideal experimental platform for frontier studies of muon physics. Using HIAF beam parameters, we design a simplified detector simulation experiment together with a supporting particle identification system to search for the $X^0$ boson whose rest mass is close to that of the muon ($m_{X^0}=105\,\text{MeV/$c^2$}$ is set in this work) and set an upper limit on its branching ratio. The GeV‑scale muon beam provided by HIAF utilizes the Lorentz boost effect to elevate low‑energy electrons from exotic decays in the muon rest frame to relativistic energies. This greatly improves the detectability of decay electrons and simplifies the detector configuration, offering a new approach to search for heavy \(X^0\) particles.

This paper is organized as follows. Section I provides an introduction. Section II describes the detector design and its functionalities for the simulation experiment. Section III presents the simulation setup and analysis procedures based on the detector model and beam samples. Section IV discusses the results obtained from the simulation studies. The final section summarizes this work and outlines prospects for future developments.

\section{Detector Design} 

\begin{figure}
    \centering
    \includegraphics[width=0.9\linewidth]{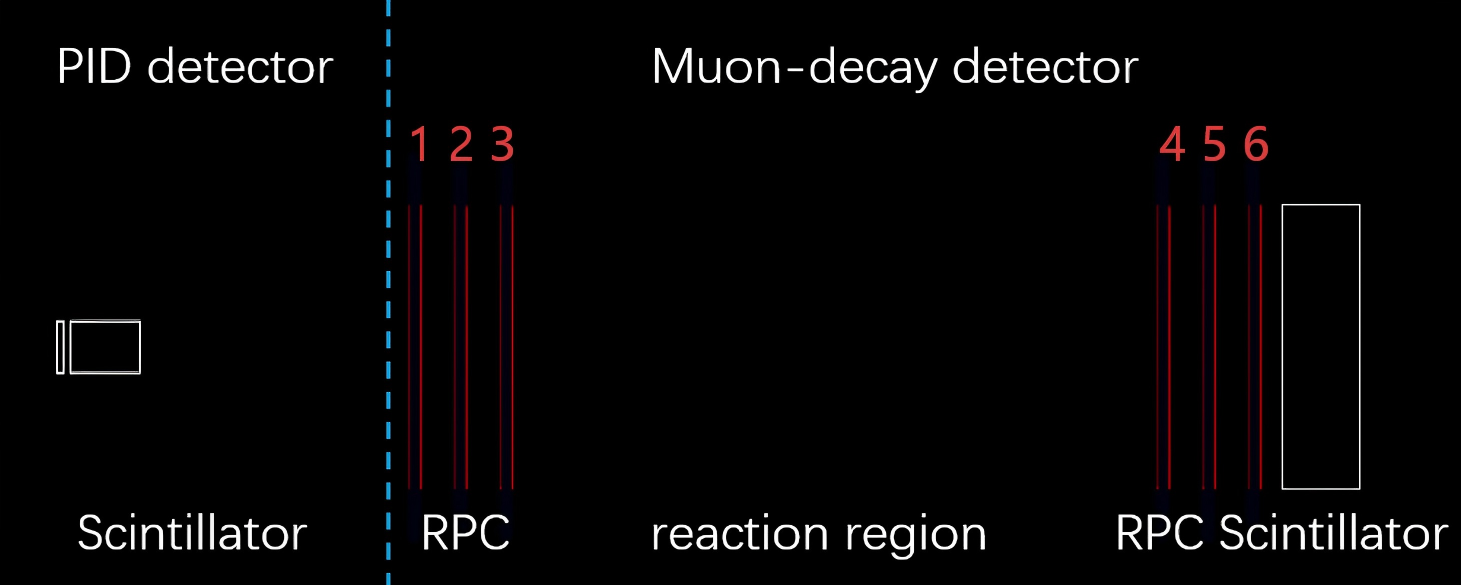}
    \caption{The detector layout, which is rendered via the visualization module of G4. The beam travels from left to right across the geometry. The PID detector occupies the left section, while the main detector system is placed on the right.}
    \label{fig:illustration_detection_system}
\end{figure}

For the simulation study presented in this work, the detector system we designed consists of two subsystems, which are responsible for PID and decay measurement, respectively. An illustration of the detection system is shown in Fig.~\ref{fig:illustration_detection_system}. All detector simulations in this work are implemented with the Geant4 (G4) toolkit\cite{GEANT4:2002zbu,Allison:2016lfl}. 

\subsection{PID System}

The muon beam from HIAF is contaminated with other charged particles, primarily pions, electrons, and protons. Since HIAF selects muons based on magnetic rigidity identification\cite{Xu:2025spd}, it is difficult to reject other particles carrying the same charge and similar momentum. To ensure the purity of the detected particle sample, a dedicated PID system for HIAF is proposed.

The PID subsystem consists of a 1-cm-thick plastic scintillator\cite{Kharzheev:2019tfk} and a 10-cm-thick cesium iodide (CsI) scintillator\cite{Adams:1995yg}. The former serves as the ToF-end detector, which works in conjunction with the ToF-start detector — a thin plastic scintillator implemented within the G4beamline simulator\cite{Roberts:2007nte}, located approximately $150\, \text{m}$ upstream — to measure the beam time-of-flight (ToF). The latter functions as an energy-deposition detector, measuring the energy deposited by beam particles inside the CsI scintillator volume. In general, particles of different species with identical kinetic energy possess different velocities (and thus different Lorentz factors). As a result, they exhibit differing time-of-flight and energy deposition signatures, which can be exploited to perform particle identification.

The simulation validation of the PID system is presented in Appendix A. The beam purity after the PID system reaches 99.96\%.

\subsection{Main Detector}

The main decay detector consists of two components: the Resistive Plate Chamber (RPC)\cite{Santonico:1981zz} stack and the CsI scintillator for energy measurement. The stack contains six RPCs with 2-D readout based on metal readout strips and delay line, numbered 1 to 6 sequentially along the beam direction. Three RPCs are arranged at the reaction volume's upstream and the others are at downstream, each three RPCs determining a straight track of the particle. When a muon decays in the reaction volume, the 6-RPC stack can provide its decay angle through the incident and exit tracks. The RPC has a  sensitive area of $28\cross28 \, \text{cm}^2$ and a thickness of $4\, \text{cm}$. The reaction volume is $90\, \text{cm}$ long, and the three RPCs installed on each side are spaced by $2.5\, \text{cm}$. This layout is designed to maximize the fraction of muons that decay within the reaction volume.

Decayed electrons and undecayed muons pass through layers 4, 5, and 6 of the RPC before entering the CsI scintillator downstream of the RPC stack, where they deposit their energy. The scintillator shares the same active area as the RPCs and has a thickness of $10\, \text{cm}$. With this thickness, the target signal electrons that lose partial energy in the downstream RPC stack deposit their full energy in the scintillator. According to the simulation, the signal electrons from the exotic decay channel, with primary momentum about $10\, \text{MeV/c}$, lose approximately 20\% of their energy before reaching the downstream CsI scintillator, while leave tracks in the downstream RPC stack, with roughly 90\% in Module 4, 5\% in Module 5, and mostly no hits in Module 6. Such performance generally meets the requirements of subsequent experiments, as our event selection is primarily based on Module 4 RPC and the downstream scintillator. Alternatively, the signal statistics could in principle be improved if the RPCs are replaced by detectors more sensitive to the energy scale of electrons from exotic decays, such as Micro-Pattern Gaseous Detectors (MPGD, e.g. GEM\cite{Sauli:1997qp}) or Scintillating Fiber Tracker (SFT\cite{Atkinson:1986ar}), for track reconstruction. More details are presented in the following.

\section{Simulation Analysis}

\subsection{Data Analysis}

In terms of simulation configuration, the beam parameters of HIAF adopted in our simulation are set as follows: beam energy of $1.2\, \text{GeV}$, beam spot diameter of $3\, \text{cm}$, and beam intensity of particles per second of $10^5-10^6$. We simulate statistical quantities corresponding to a continuous 1-month data-taking period, which yields approximately $10^{12}$ total raw muon events. Given the muon half-life $\tau_0=2.2\cross 10^{-6}\,\text{s}$\cite{ParticleDataGroup:2024cfk}, the decay probability of muons at the corresponding energy under relativistic effects is expressed as 

\begin{equation}
    P_{\text{decay}}\left(l\right)=1-\exp\left(-\frac{l}{\gamma c \tau_0}\right),
    \label{eq2}
\end{equation}

where $l$ denotes the flight distance of the muon and $\gamma=E_{\mu}/m_{\mu}$. For muons that decay within the reaction volume about $1\, \text{m}$, the decay probability is on the order of $10^{-4}$, corresponding to roughly $10^8$ decay events. For muon events with potential decays, we weight the decay probabilities of muons at different positions inside the reaction volume according to Eq.~\ref{eq2}, so that the simulation better reproduces the actual physical scenario. During the preliminary study, the branching ratio of exotic decays was fixed at 0.01\% to perform comparisons between the contributions of the signal and the background.

In this experiment, the signal of interest refers to electrons produced via exotic decay channels \(\mu\to eX^0\), where $m_{X^0}=105\,\text{MeV/$c^2$}$. The background sources fall into two major categories: the first originates from undecayed muons passing directly through the detector, while the second consists of electrons generated from SM decay channels. 

\begin{figure}
    \centering
    \includegraphics[width=0.9\linewidth]{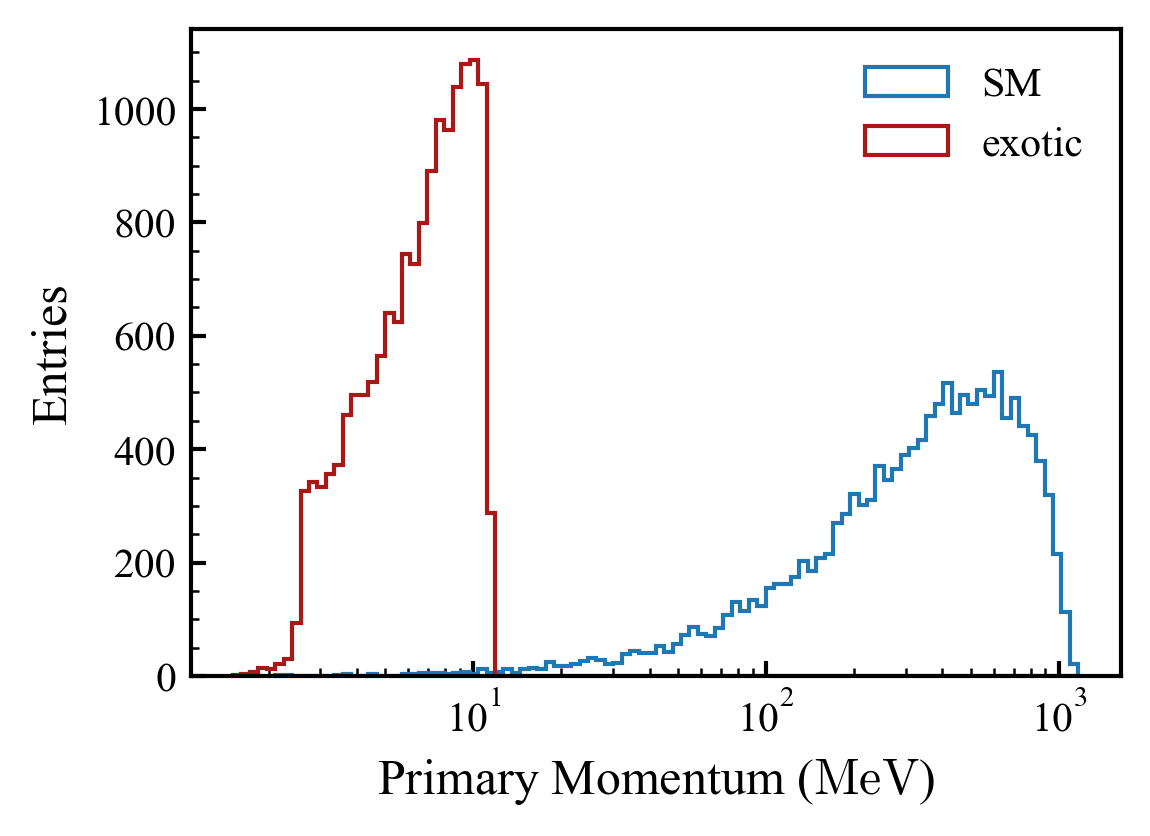}
    \caption{primary momentum distributions of decay electrons obtained from simulations. Evidently, the initial‑momentum spectra of decay electrons from the two models exhibit clear differences. It should be noted that the electrons from the exotic‑channel decay show a uniform distribution when the horizontal axis is plotted in linear scale.}
    \label{fig:primary_momentum}
\end{figure}

We first compare the primary momentum spectra of electrons from the two decay modes, which is shown in Fig.~\ref{fig:primary_momentum}. Electrons from SM decays exhibit a broad momentum distribution. In contrast, since the mass of the $X^0$ is close to the muon mass, only a small amount of kinetic energy is released during exotic decays. Consequently, primary electron momenta are highly concentrated around $10\, \text{MeV/c}$. Electrons within this energy range deposit their full energy in the downstream scintillator, enabling discrimination against higher-energy undecayed muon background and most background electrons originating from SM decays. This distinct momentum feature therefore serves as the primary handle to separate signal from background. 

We further utilize angular distribution features to suppress background. We adopted the assumption of an unpolarized muon beam throughout all simulations. The exotic channel corresponds to a simple two-body decay. According to Eq.~\ref{eq1}, the decay electron carries an energy $E_e^*=0.66\, \text{MeV}$ in the muon rest frame and follows a monoenergetic isotropic distribution ($d\Gamma'/d\cos\theta'=1/2$). In the laboratory frame, if the differential decay width for muon decays via the exotic channel also follows\cite{Okun:1984pq}:

\begin{align}
    d\Gamma_{\text{PW}} = \frac{\langle |\mathcal{M}|^2 \rangle}{2E_\mu} (2\pi)^4 \delta^{(4)}(p_\mu-p_e-p_{X^0}) \nonumber \\ \times \frac{\mathrm{d}^3p_e}{(2\pi)^3 2E_e} \frac{\mathrm{d}^3p_{X^0}}{(2\pi)^3 2E_{X^0}},
    \label{eq3}
\end{align}

where \(\mathcal{M}\) denotes the decay amplitude, and $p_i$ represents the four-momentum of the corresponding particle. Thus, integrating over the final-state phase space and imposing energy–momentum conservation yields\cite{GUO:2026jua}:

\begin{equation}
    \frac{\mathrm{d}\Gamma_{\text{PW}}}{\mathrm{d}E_{e}} = \frac{\langle |\mathcal{M}|^2 \rangle}{16\pi E_{\mu} \sqrt{E_{\mu}^2 - m^2}},
    \label{eq4}
\end{equation}

\begin{equation}
    \frac{\mathrm{d}\Gamma_{\text{PW}}}{\mathrm{d}\cos\theta} = \frac{\langle |\mathcal{M}|^2 \rangle}{16\pi E_{\mu}}\sum_i \frac{\sqrt{E_{e,i}^2(E_{\mu},\theta) - m_{e}^2}}{E_{\mu} - E_{e,i}(E,\theta)},
\label{eq5}
\end{equation}

with

\begin{equation}
    \cos\theta = \frac{2E_{\mu}E_{e} - m_{\mu}^{2} - m_{e}^{2} + m_{{X^0}}^{2}}{2\sqrt{\left(E_{\mu}^{2}-m_{\mu}^{2}\right)\left(E_{e}^{2}-m_{e}^{2}\right)}}.
    \label{eq6}
\end{equation}

Eq.~\ref{eq4} corresponds to the uniform distribution shown as the red histogram in Fig.~\ref{fig:primary_momentum}, and the energy boundaries calculated by substituting numerical values show good consistency. According to Eq.~\ref{eq6}, a smaller \(E_e^*\) imposes an upper limit on the decay angle, such that a single \(\theta\) corresponds to two different values of \(E_e\). This accounts for the summation term in Eq.~\ref{eq5}.

\begin{figure}[htb]
  \centering
  \includegraphics[width=1\linewidth]{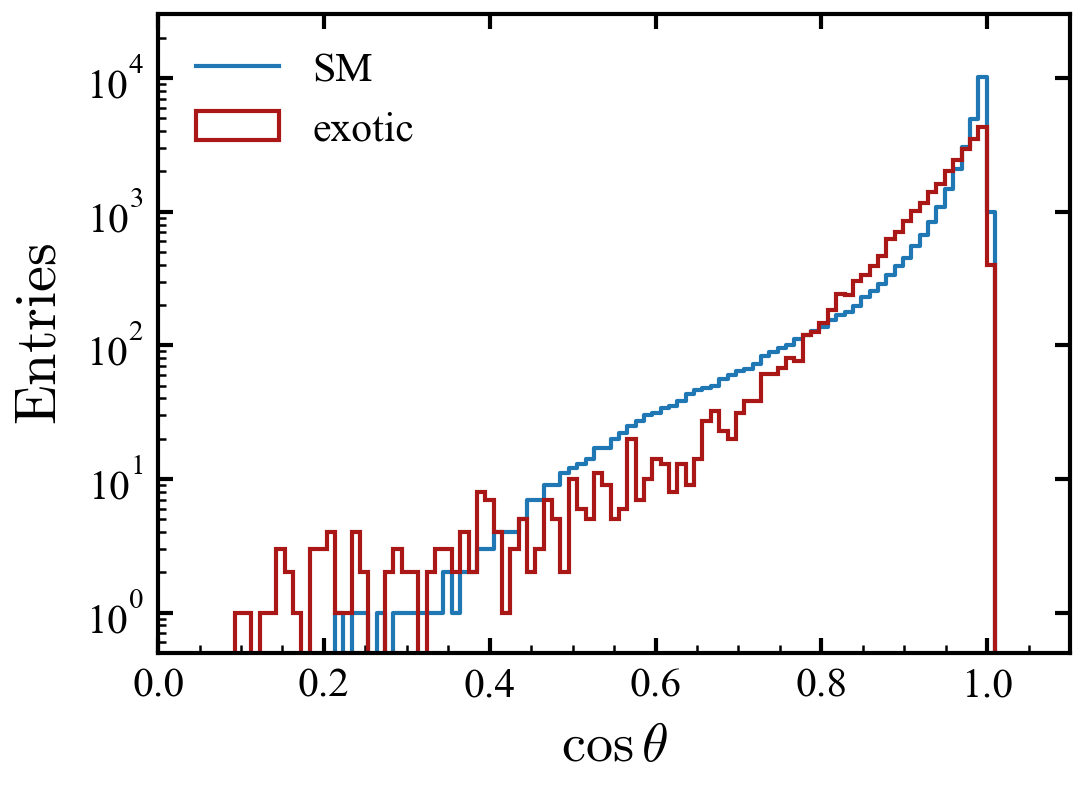}
  \vspace{6mm}
  \includegraphics[width=1\linewidth]{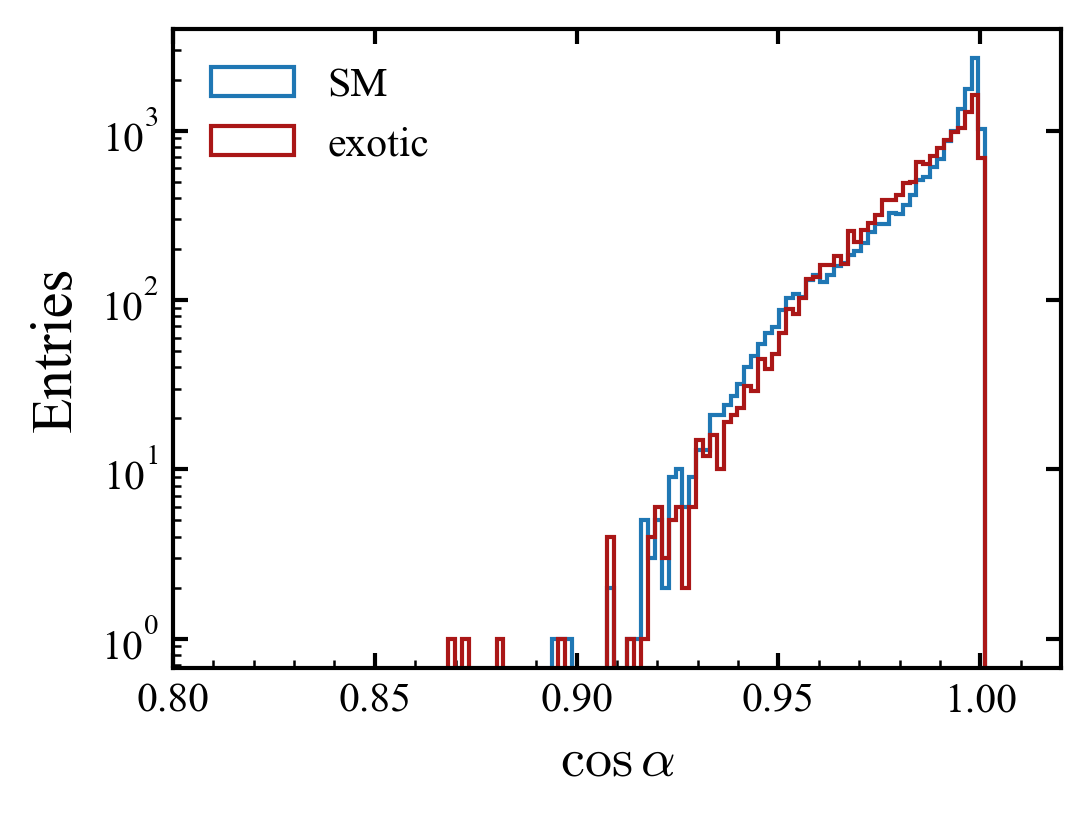}
  \caption{Upper panel: The decay-electron angular distribution is computed via the POCA algorithm using the hit-point coordinates of simulated data on the RPCs. Lower panel: Distribution of decay-electron hit positions on layer 4 of the RPC, expressed in terms of angle.}
  \label{fig:ang_distribution}
\end{figure}

The decay angular distribution in the SM (three-body decay) in the muon rest frame is given by the Michel distribution\cite{Michel:1949qe}, which can be Lorentz-boosted to the laboratory frame. Although the decay electrons in the muon rest frame are no longer monoenergetic, the vast majority possess higher energies than those from exotic decay channels. Upon Lorentz boosting to the laboratory frame, they consequently occupy a broader angular distribution, as verified in Fig.~\ref{fig:ang_distribution}.


The upper panel of Fig.~\ref{fig:ang_distribution} presents the angular distributions extracted from the simulated decays using the Point of Closest Approach (POCA) algorithm, which geometrically finds the minimal-distance points on two extrapolated tracks (generally non-coplanar) to reconstruct the decay vertex and emission angle. Considering that the target electrons from exotic decays can hardly be recorded by layers 5 and 6 of the RPC, the data used in our calculations are read directly from G4. Scattering of particles in air and detector materials introduces discrepancies between the decay angular distribution reconstructed via the POCA algorithm and theoretical predictions. Nevertheless, the overall trend persists: the SM background exhibits a broader angular distribution. This trend is visualized by the hit positions of decay electrons on layer 4 of the RPC, as shown in the lower panel of Fig.~\ref{fig:ang_distribution}, where \(\alpha=\arctan (2r/L)\) denotes the angle between the line connecting the hit point and the center of the reaction region and the central axis of the detector system, $r$ is the distance from the hit point to the center of the RPC, and $L$ is the length of the reaction region. Accordingly, we select a limited geometrical area on layer 4 of the RPC as the angular cut in our simulation.

Besides the two categories of backgrounds introduced at the beginning of this chapter, namely Standard Model muon decays and muon-related backgrounds, additional background sources can contribute to the total background in the full simulation. A dominant one comes from the decay channel \(\mu \to eee\nu\nu\). In this process, electrons produced via internal conversion \(\gamma \to e^+e^-\) can acquire nearly light-like boost factors. As a result, their energies in the muon rest frame may approach that of the signal electrons and constitute an irreducible background. We performed dedicated G4 simulations for this decay channel. A total of 15600 generated events were simulated, and 651 events survived the selection within the target momentum window of 2 to $12\, \text{MeV/c}$. Taking into account its branching ratio of \(3.4\times 10^{-5}\), the expected number of background events under our simulation conditions is approximately 142. Compared with the 8131 signal events after the same momentum selection, this background induces a bias of roughly 1.7\%. The resulting upper limit shifts by about 1.7\% accordingly, which we consider negligible within the statistical uncertainty.

We also discuss other possible background sources. Cosmic muons have a flux more than five orders of magnitude below the beam flux and can be safely ignored. Pions are largely rejected by PID, and only a small fraction decays with essentially no effect on the result. With the $1\,\text{ns}$ RPC time resolution, the pileup probability is negligible (\(10^{-7}\)). Ambient ionizing electrons are removed by the front-end veto because of geometric acceptance. Knockout electrons from muon-air interactions can be eliminated using a vacuum environment. And in terms of the intrinsic noise originating from Module 4 RPC, after suppression and cluster selection, it is barely recorded by the downstream CsI scintillator detector.

\section{Result} 

\begin{figure}
    \centering
    \includegraphics[width=1\linewidth]{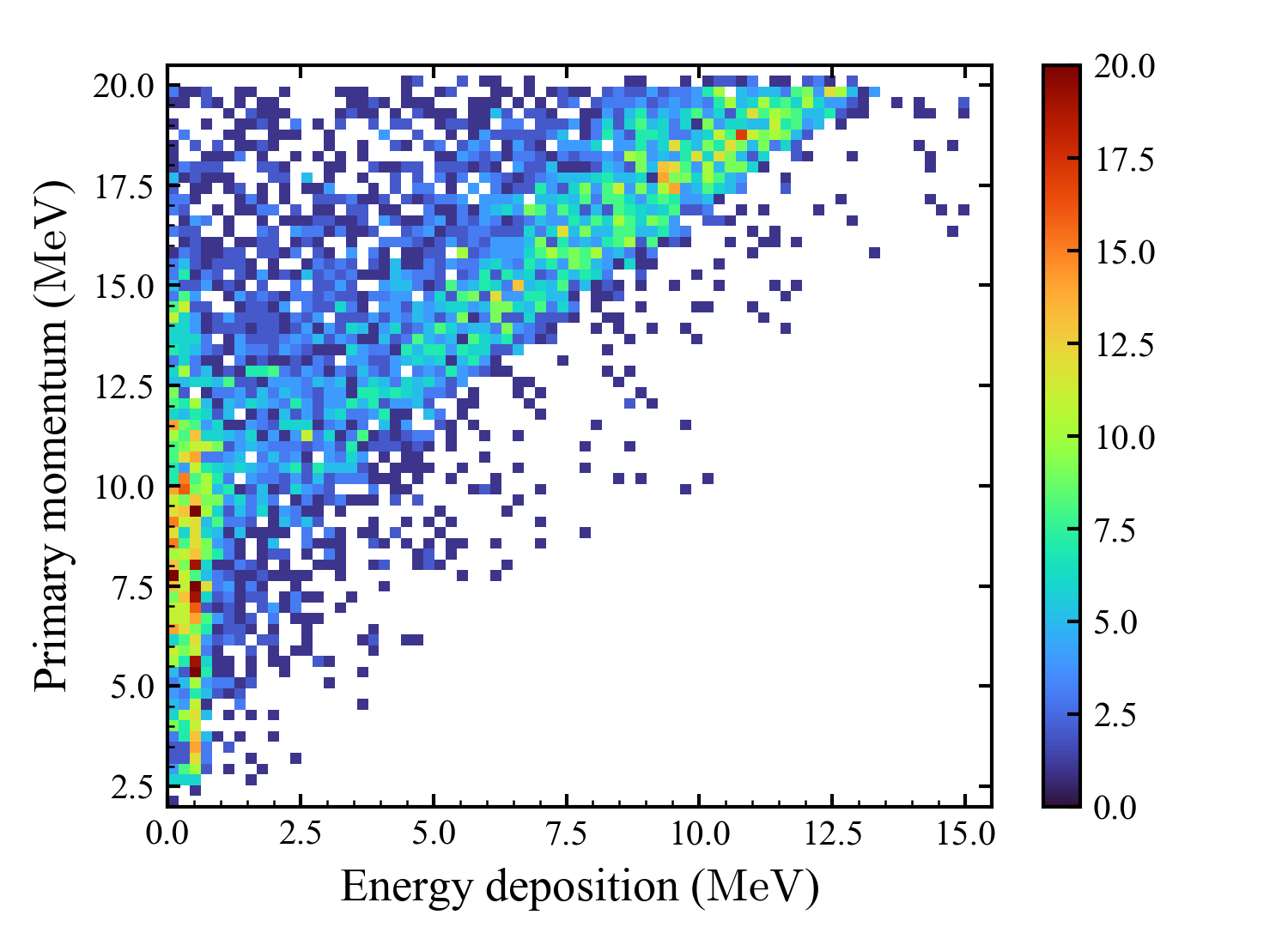}
    \caption{Correlation between the primary momentum of simulated decay electrons and their energy deposition in the downstream scintillator. For deposited energies above $2\, \text{MeV}$, a linear correlation between primary momentum and deposited energy can be observed. The intercept of this linear relation corresponds to the energy loss of electrons in the downstream RPC stack.}
    \label{fig:momentum_Edeposition}
\end{figure}

Before defining the analysis selection cuts, we first account for the effect of the detector system on the energy of the target electrons. The low-energy electrons produced in decays pass through three layers of RPC detectors downstream of the reaction volume before reaching the downstream CsI crystals, losing a fraction of their kinetic energy during traversal. The primary momentum versus deposited energy spectrum of electrons is presented in Fig.~\ref{fig:momentum_Edeposition}. The simulation results indicate that the average energy loss in these RPC layers is approximately 7 to $9\, \text{MeV}$. According to simulation data, electrons with primary momenta of 2 to $12\, \text{MeV/c}$ deposit energy ranging roughly from 0 to $3\, \text{MeV}$ in downstream CsI crystals, with the vast majority of such energy depositions concentrated within the 0 to $1\, \text{MeV}$.

\begin{figure}
    \centering
    \includegraphics[width=1\linewidth]{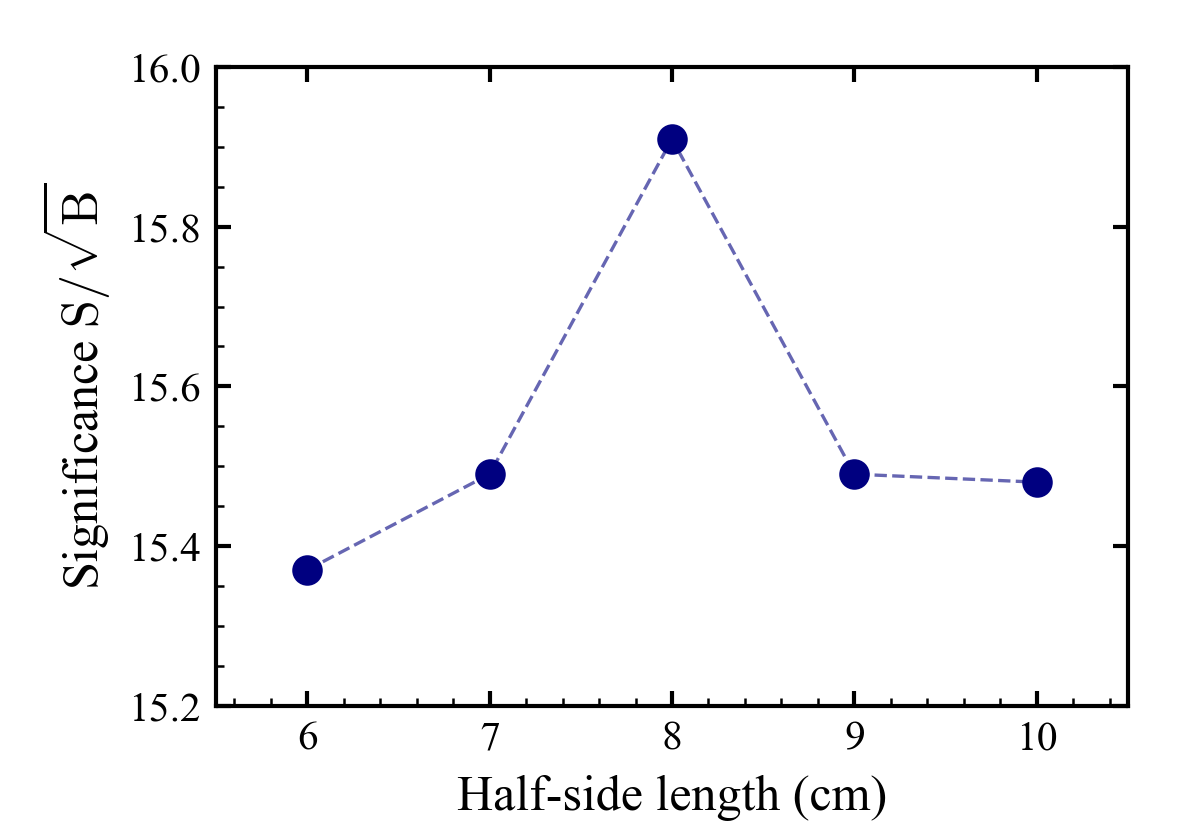}
    \caption{Variation of significance as a function of the half‑side length of the square selection region defined on Module 4 RPC. The optimal significance is achieved for a half‑side length of $8\, \text{cm}$.}
    \label{fig:significance_vs_cut}
\end{figure}

In the final event selection, we mainly consider two variables: electron angle and energy. Here, the angular cut refers to the spatial bound imposed on layer 4 of the RPC, while the energy cut refers to the restriction on the energy deposited in the downstream scintillator. Given the relatively narrow energy window, we perform angular cuts first, and subsequently optimize the energy cut under the optimal angular selection. As discussed above, an angular acceptance cut defined by the RPC detector geometry is implemented in the event selection. Specifically, a set of square-shaped fiducial regions including half‑side lengths from 6 to $10\, \text{cm}$ is defined on Module 4 RPC plane to perform the angular cuts. We count the number of signal and background events for each fiducial region with the energy cut on the primary momentum range of 2 to $12\, \text{MeV/c}$ and evaluate the statistical significance using $\text{S}/\sqrt{\text{B}}$. The results are summarized in Fig.~\ref{fig:significance_vs_cut}. Although significance varies slightly between different selections, the square region with a half-side length shorter than $8\,\text{cm}$ is adopted as the final angular cut.

Then, we adopted two energy deposition cuts: 0 to $1\, \text{MeV}$ and 0 to $3\, \text{MeV}$. The primary momentum range of 2 to $12\, \text{MeV/c}$ was used for comparison. We use CMS Combine\cite{CMS:2024onh} to calculate the significance and aim to derive an exclusion limit on the decay branching ratio based on simulation studies. The expected number of events satisfies

\begin{equation}
    N(r,\theta) = r S_{0} + B(\theta),
    \label{eq7}
\end{equation}

where $S_0$ denotes the signal yield evaluated at the reference branching ratio \(\mathrm{BR}_\mathrm{ref}\), $B(\theta)$ the background yield, \(\theta\) the nuisance parameter for systematic uncertainties, and $r$ the parameter of interest. The physical branching ratio is given by \(\mathrm{BR}=r\,\mathrm{BR}_\mathrm{ref}\). We perform a maximum-likelihood fit to the Poisson likelihood. The 95\% confidence-level upper limit is extracted as the value of $r$ satisfying \(\mathrm{CL}_s = \mathrm{CL}_{s+b}/\mathrm{CL}_b = 0.05\), and the upper limit branching ratio \(\text{BR}_\text{limit}=r\text{BR}_\text{ref}\).

In principle, reducing the beam energy in simulation shortens the laboratory-frame muon lifetime and increases the particle time-of-flight. This allows more muons to decay within the reaction region, enlarging the available event sample and improving the statistical significance. Nevertheless, such a scheme would also lower the energy of decay electrons. Taking into account the counting efficiency of the RPC stack and the downstream CsI scintillator for low-energy electrons as well as the associated energy loss during penetration\cite{Jamil:2010zz}, we ultimately opted against reducing the beam energy. 

 The event yields obtained under different selection cuts are summarized in Table.~\ref{tab2}, with the angular restriction corresponding to a half-side length smaller than $8\, \text{cm}$. Meanwhile, we present the upper limit of the confidence-level of 95\% on the branching ratio \(\text{BR}_\text{limit}=r\text{BR}_\text{ref}\simeq10^{-5}\), obtained using the CMS Combine framework with the corresponding datasets, where $\text{BR}_\text{ref}=0.01\%$ . This demonstrates that our detector design and event selection strategy are reasonable and feasible, and the setup holds the potential to carry out an experimental search for this CLFV process. Considering the detector uncertainty, which is estimated at 1\%, the resulting systematic uncertainty on the surviving counts after the energy cut in Table.~\ref{tab2} is approximately $0.01\times347224/6240=55.6\%$. This large uncertainty arises from the modest detector resolution and substantial background level, and can influence the upper limit as $0.16_{-0.06}^{+0.07}$. Future measurements with higher-resolution detectors would enable tighter selection criteria and thus substantially reduce this systematic uncertainty.
 
 \begin{table}[h]
  \begin{ruledtabular}
    \begin{tabular}{cccccccccc}
      Cut & Signal & Background & Limit $r$\footnotemark[1] \\ \hline
      0 to $1\, \text{MeV}$ & 4875 & 280311 & 0.21 \\ \hline
      0 to $3\, \text{MeV}$ & 6240 & 347224 & 0.16 \\ \hline
      2 to $12\, \text{MeV/c}$\footnotemark[2] & 8131 & 261109 & 0.12 \\
    \end{tabular}
  \end{ruledtabular}
  \caption{Simulated events and upper limit of branching ratio ($\text{BR}_\text{limit}$) under different cut with $\text{BR}_\text{ref}=0.01\%$. 2 to $12\, \text{MeV/c}$ refers to the primary momentum cut, the others refer to the energy deposition cuts.}
  \label{tab2}
  \footnotetext[1]{$r=\text{BR}_\text{limit}/\text{BR}_\text{ref}$}
  \footnotetext[2]{The primary momentum information is accessible only in simulations and is shown for reference.}
\end{table}

\section{Summary and Outlook} 

This work proposes a complete experimental search scheme for the lepton-flavor-violating process $\mu \to eX^0$ based on the PKMu platform\cite{Ruzi:2023mxp} and the HIAF muon beamline. The detector design and simulation validation have been fully completed.

We have designed a PID system combining time-of-flight and deposited-energy measurements. Simulations verify its particle identification capability for 1 GeV-scale beams, which can efficiently separate muons and pions with similar masses and meet the experimental requirements. Taking advantage of the high Lorentz factor of the HIAF muon beam, we successfully boost the low-energy electrons produced in decays involving the $105\, \text{MeV/$c^2$}$ \(X^0\) particle into a detectable energy range. Exploiting the low momentum and small emission angle of electrons originating from the exotic decay channel, we efficiently suppress backgrounds by applying momentum and angular acceptance cuts. Finally, an upper limit of $10^{-5}$ on the decay branching ratio is obtained at the 95\% confidence level, which is marked by the star symbol in Fig.~\ref{fig:introduction_results} and corresponds to world-leading sensitivity. Compared with other results shown in Fig.~\ref{fig:introduction_results}, our result relies on high‑energy muon beams, which may open a new path for future similar studies. Meanwhile, it can be anticipated that for \(X^0\) masses closer to the muon mass, the initial‑momentum range of decay electrons will be further compressed. Corresponding cuts can reject more background events, which allows the branching ratio upper limit to be further improved.

Naturally, due to the muon beam relativistic effects and the detector’s finite energy resolution, the exotic muon decay signal appears only as a smooth low-energy electron excess above the Standard Model background, with no distinct peak in the electron spectrum. Our analysis is therefore an excess search rather than a peak search. The \(\mu\to eX^0\) decay is a compelling interpretation of our result.

At the present stage, this study remains in the simulation phase. In subsequent work, we will further optimize the detector geometry and develop improved event selection algorithms to enhance the detection sensitivity. Meanwhile, more detailed simulations of detector response and comprehensive analyzes of all background sources will be performed to bring the simulation results closer to real experimental conditions. Looking ahead, we plan to formally construct the detector apparatus at HIAF and conduct data taking to experimentally validate the design presented in this work.

\section{Acknowledgment}

This work is supported in part by the National Natural Science Foundation of China under Grants No. 12325504, and No. 12522507.


\appendix

\section{Appendix A: PID Simulation} 

\begin{figure}
    \centering
    \includegraphics[width=0.9\linewidth]{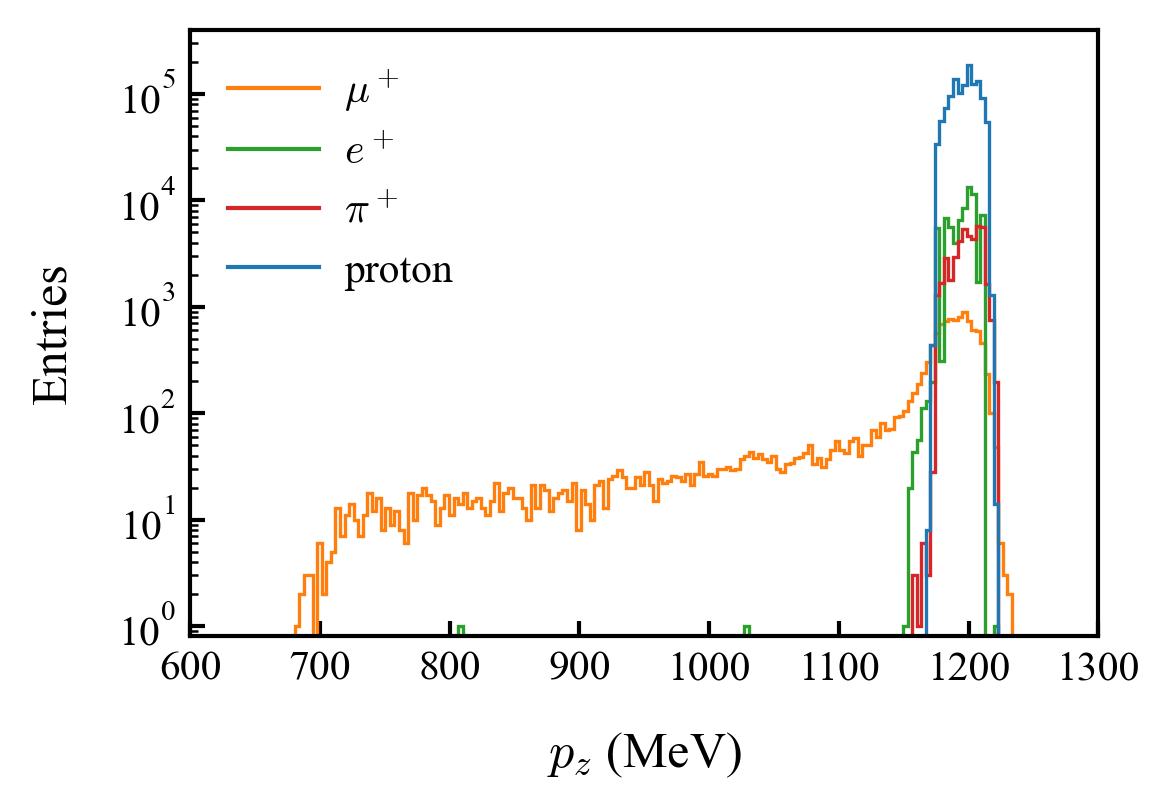}
    \caption{The composition of the simulated beam used to test PID performance. The primary generator is implemented via the G4beamline simulator. All particles are generated with an energy of $1.2 \, \text{GeV}$, and an additional low-energy tail is included for the muon component.}
    \label{fig:PID_beam_composition}
\end{figure}

\begin{figure}[htb]
  \centering
  \includegraphics[width=1\linewidth]{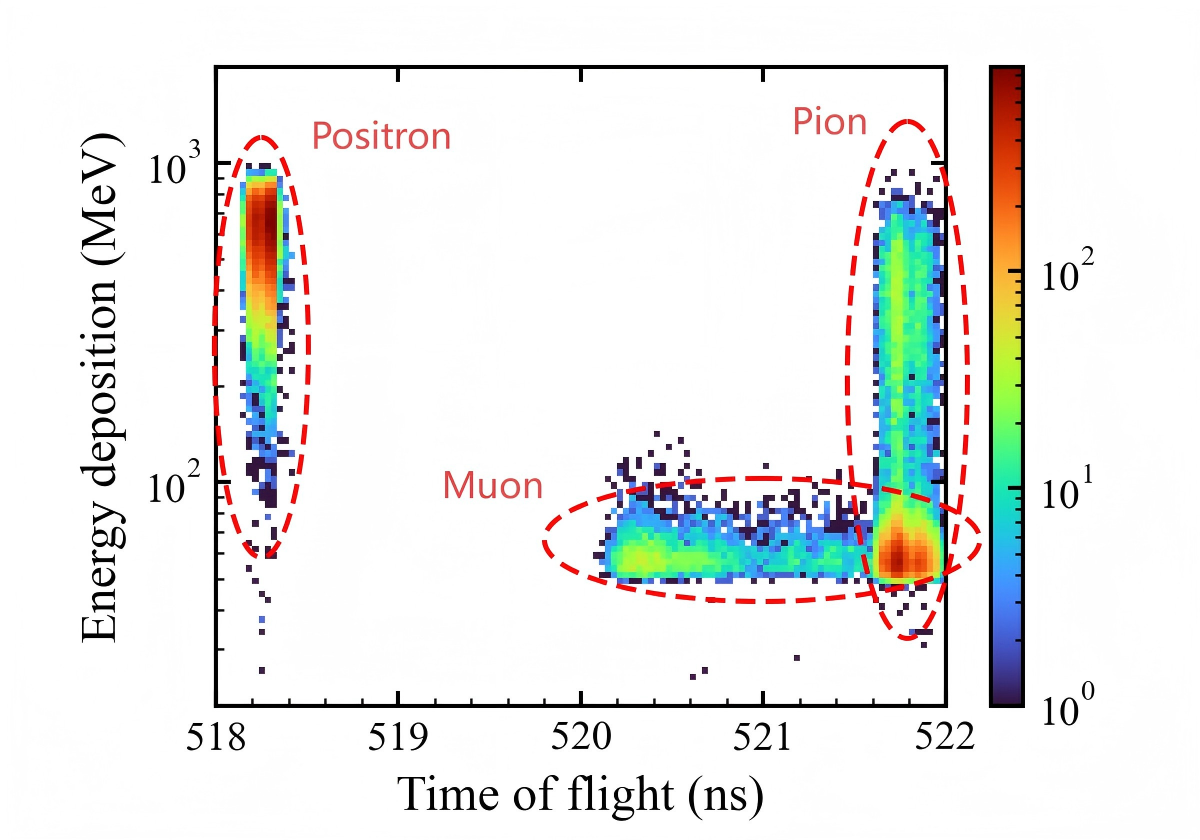}
  \vspace{6mm}
  \includegraphics[width=1\linewidth]{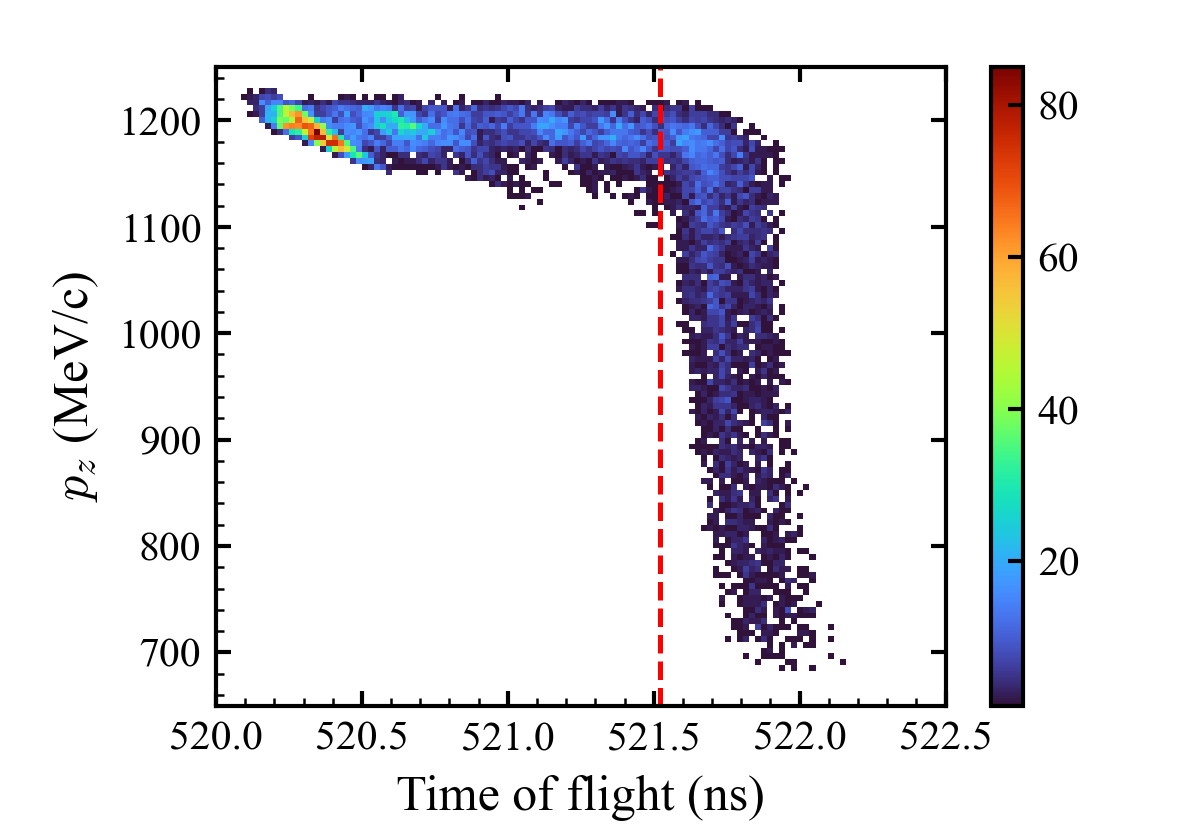}
  \caption{Simulation results of the PID detector. Upper panel shows the correlation between deposited energy and time-of-flight, and lower panel presents the longitudinal momentum versus time-of-flight distribution. Distributions corresponding to different particle species are labeled in each plot.}
  \label{fig:PID_simulation}
\end{figure}

We perform simulation-based particle identification for the $1.2 \, \text{GeV}$ beam, and the beam composition is illustrated in Fig.~\ref{fig:PID_beam_composition}. We recorded the TOF and deposited energy information, which are presented in Fig.~\ref{fig:PID_simulation}. Protons are not shown in this plot, as their flight time differs drastically from other particle species. The results show that positrons ($e^+$) can be easily discriminated via their deposited energy. In contrast, since the masses of positive muons ($\mu^+$) and positive pions ($\pi^+$) are close to each other, and $\mu^+$ particles originate from decays of $\pi^+$, their distributions overlap in the spectrum, making them comparatively difficult to separate.

We further examine the longitudinal momentum $p_z$ of particles falling within the TOF range of muons, as illustrated in Fig.~\ref{fig:PID_simulation}. The analysis shows that the overlap region between $\mu^+$ and $\pi^+$ corresponds exactly to the low-$p_z$ interval of $\mu^+$. Consequently, we can impose a suitable threshold on the TOF variable to simultaneously reject $\pi^+$ and low-momentum muons, achieving excellent particle identification performance.

\bibliographystyle{unsrt}
\bibliography{ref.bib}
\end{document}